# Structural and WAL analysis of Topological single-crystal SnSb$_2$Te$_4$

Ankush Saxena[1,2], M.M. Sharma[1,2], Prince Sharma[1,2], Yogesh Kumar[1,2], Poonam Rani[2,4], M. Singh[3], S. Patnaik[3,] and V.P.S. Awana[1,2*]

[1]*Academy of Scientific & Innovative Research (AcSIR), Ghaziabad-201002, India*
[2]*CSIR- National Physical Laboratory, New Delhi-110012, India*
[3]*School of Physical Sciences, Jawaharlal Nehru University, New Delhi-110067, India*
[4]*Materials Science Division, Inter-University Accelerator Centre, New Delhi-110067, India*

**Abstract:**

Here, we report successful single crystal growth of SnSb$_2$Te$_4$ using the self-flux method. Unidirectional crystal growth is confirmed through X-ray Diffraction (XRD) pattern taken on mechanically cleaved crystal flake while the rietveld refined Powder XRD (PXRD) pattern confirms the phase purity of the grown crystal. Scanning Electron Microscopy (SEM) image and Energy Dispersive X-Ray analysis (EDAX) confirm crystalline morphology and exact stoichiometry of constituent elements. Vibrational Modes observed in Raman spectra also confirm the formation of the SnSb$_2$Te$_4$ phase. DC resistivity ($\rho$-T) measurements confirm the metallic character of the grown crystal. Magneto-transport measurements up to ±5T show a non-saturating low magneto-resistance percentage (MR%). V-type cusp and Hikami Larkin Nagaoka (HLN) fitting at lower field confirms the Weak Anti-localization (WAL) effect in SnSb$_2$Te$_4$. Density Functional Theory (DFT) calculations were showing topological non-trivial electronic band structure. It is the first-ever report on MR study and WAL analysis of SnSb$_2$Te$_4$ single crystal.

**Keywords**: Topological Insulators, Crystal Growth, Raman Spectroscopy, Magneto-resistance, Weak Anti-localization effect, Density Functional Theory.

PACS: 74.70.Dd, 74.62.Fj, 74.25.F-

*Corresponding Author

Dr. V. P. S. Awana: E-mail: awana@nplindia.org

Ph. +91-11-45609357, Fax-+91-11-45609310

Homepage: awanavps.webs.com

**Introduction:**

The discovery of Topological materials has revolutionized the field of condensed matter physics. Ever since the unearthing of topological materials, condensed matter scientists are always keen to search for new materials of such kind. Topological Insulator (TI) is the most studied class of topological materials [1-4]. Topological insulators (TIs) are characterized as materials that have a fully insulating bulk along with the conducting surface states (SS) [2]. These SS are formed due to intrinsic spin-orbit coupling (SOC) in TIs [2,3]. These surface states are robust in nature as these are protected by Time-Reversal Symmetry (TRS) [1]. The presence of time-reversal symmetry in TIs generates doubly degenerate SS, which have opposite spins; these are known as Kramer's doublet [2,5]. The spins of carriers in these surface states are locked transversely to their momentum, and this phenomenon is known as spin-momentum locking [2,3]. This spin momentum locking is evidenced by observing the weak anti-localization (WAL) effect in magneto-transport measurements of TIs [6,7]. The presence of SS makes TIs very crucial materials to observe various phenomena such as topological superconductivity [8,9], high magnetoresistance [10,11], Dirac fermions [12,13], and many more. The presence of robust surface states and intrinsic SOC make TIs quite fruitful in the field of spintronic [4].

Till now, the most studied TIs include $Bi_2Se_3$, $Bi_2Te_3$, $Sb_2Te_3$; they contain a single Dirac cone on their surface [14]. Very recently discovery of a new magnetic topological insulator $MnBi_2Te_4$ [15,16] created a new venture for TIs. In these new TIs, layers of different compounds are introduced in the lattice of parent TI as in $MnBi_2Te_4$, the layer of MnTe is inserted in the lattice of $Bi_2Te_3$ [15]. Following this, some other TIs have been discovered such as $FeBi_2Te_4$ [17], $SnBi_2Te_4$ [18,19], $PbBi_2Te_4$ [18,19]. Among these, $FeBi_2Te_4$ and $MnBi_2Te_4$ come under the category of magnetic TIs [15,17]. $SnBi_2Te_4$ and $PbBi_2Te_4$ are conventional TIs, and these can be symbolically written as $A_{IV}$-$B_{VI}$-$(A_V$-$B_{VI})_m$, here m=1,2,3.. where subscript represent the respective groups of elements in the periodic table [18]. Another member of this family is $SnSb_2Te_4$, in which a layer of SnTe is inserted into the $Sb_2Te_3$ lattice [20]. $SnBi_2Te_4$ and $PbBi_2Te_4$ are theoretically predicted to have topological non-trivial character [18,21]. The presence of surface states has been experimentally visualized in $PbBi_2Te_4$ through Angle-resolved spectroscopy (ARPES) measurements [22,23], but this type of confirmation is yet to be done for $SnBi_2Te_4$. Also, the existence of topological surface states in $SnBi_2Te_4$ and $PbBi_2Te_4$ single crystals has been confirmed by SdH oscillations, and the low field weak anti-localization effect has also observed [24]. There are very few reports on the synthesis of other compounds of this series, i.e., $SnSb_2Te_4$ [25,26]. $SnSb_2Te_4$ is found to have better surface states as compared to parent compound $Sb_2Te_3$ in DFT calculations [21,27], but their experimental realization through ARPES measurements is still missing. $SnSb_2Te_4$ is found to show pressure-induced superconductivity in high-pressure transport measurements [28], whereas a transition from diffusive to hopping transport has been observed in $SnSb_2Te_4$ thin films. The presence of a weak anti-localization effect causes the negative magneto-conductivity in a perpendicular magnetic field for metallic samples [29]. Yet, there is no report available on magneto-transport measurements of single-crystalline $SnSb_2Te_4$.

In this article, we report the successful growth of single-crystalline $SnSb_2Te_4$ through the self-flux method. XRD pattern taken on crystal flake confirms the crystallinity of the synthesized $SnSb_2Te_4$ single crystal. Homogenous distribution of constituent elements in the stoichiometric ratio is confirmed through EDAX measurements which signifies the purity of the sample. A non-saturating but low MR has been observed in magnetotransport measurements performed at 5K. A V-like cusp below 1T in MR% shows a possible WAL effect in $SnSb_2Te_4$, which is also checked by HLN fitting. DFT-based calculations on Density of States (DOS) show non-vanishing DOS at the Fermi level. SOC is found to be effective on bulk electronic band structure in DFT-based band structure calculations. Here, it is worth mentioning that this is the first-ever report on magneto-transport measurements and the WAL effect in $SnSb_2Te_4$. This report on the WAL effect in $SnSb_2Te_4$ can be regarded as the first experimental report on the topological character of $SnSb_2Te_4$, which signifies the presence of surface states in the same.

**Experimental:**

Single crystal of $SnSb_2Te_4$ was grown by using a self-flux method by following well-optimized heat treatment. High quality (>4N) powders of Sn, Sb, and Te were taken in stoichiometric amounts. These powders were mixed and grounded by using agate mortar pestle to get a homogenous mixture. This mixture was then palletized and vacuum encapsulated in quartz ampoules at a pressure of $5\times10^{-5}$ mbar. This vacuum encapsulated sample was then heated to $890^0C$ at a rate of $120^0C/h$ in a programming-controlled muffle furnace. The sample was kept at this temperature for 4 hours so that the melt becomes homogenous. Then this melted sample was cooled down to $500^0C$ at a rate of $1^0C/h$; during this step, crystal growth occurs. After this, the sample is sintered at $500^0C$ for 150 hours and then allowed to cool generally to room temperature. The schematic of this heat treatment is shown in fig.1. Thus obtained crystal is silvery shiny and easily cleavable by using a surgical blade along the growth axis. The image of synthesized $SnSb_2Te_4$ single crystal is shown in the inset of fig.1(a).

Rigaku mini flex-II tabletop X-ray diffractometer equipped with Cu $K_\alpha$ radiation of 1.5418Å wavelength was used to record the XRD pattern of crystal flake and gently crushed powder of synthesized $SnSb_2Te_4$ single crystal. Rietveld refinement of the PXRD pattern was performed using Full Proof software, and the Unit cell of synthesized $SnSb_2Te_4$ single crystal was drawn by using VESTA software. Joel JSM 7200F FESEM is used to record SEM images and EDAX measurements. Raman Spectra is recorded by using Renishaw inVia Reflex Raman Microscope equipped with a Laser of 514nm. & 720nm. The sample is irradiated with a LASER having a wavelength of 514nm. The sample is exposed to LASER for a period of 30 sec. and the power was maintained below 5mW to avoid any local heating to the sample due to LASER. The magneto-transport studies were carried out by using the conventional four-probe method on Quantum Design Physical Property Measurement System (PPMS) equipped with a sample rotator and closed-cycle based cryogen-free system. Magneto-transport measurements were performed at a temperature of 5 K for an applied magnetic field range of ±5 T. DFT calculations were performed on Quantum Espresso software.

**Results & Discussion:**

Fig. 2(a) depicts the XRD pattern taken on mechanically cleaved crystal flake of synthesized $SnSb_2Te_4$ single crystal. This XRD pattern shows very sharp high-intensity peaks in the (003n) direction. It is typical behavior that a single crystalline material shows. It confirms that the sample has grown only along c-direction. Here it is worth mentioning that these XRD peaks are different from those observed for magnetic TIs of the same kind, such as $MnBi_2Te_4$ and $FeBi_2Te_4$. In these magnetic TIs, XRD peaks were obtained in the (004n) direction [16,17], while in $SnSb_2Te_4$, these have occurred in the (003n) direction. The reason for this different behavior lies in the unit cell of these materials. The unit cell of $MnBi_2Te_4$ and $FeBi_2Te_4$ contains 4 blocks of septuple layers, as seen in ref. 16,17. Conversely, the unit cell of $SnSb_2Te_4$ contains 3 blocks of septuple layers. Thus high-intensity XRD peaks are observed in the (003n) direction.

Fig. 2(b) depicts Rietveld's refined PXRD pattern of synthesized $SnSb_2Te_4$ single crystal. Rietveld refinement confirms that the sample is crystallized in rhombohedral crystal structure with R -3 m space group symmetry. No impurity peak can be seen in the Rietveld refined PXRD pattern of synthesized $SnSb_2Te_4$ single crystal, and this confirms that the sample has grown in a single phase. The quality of fit ($\chi 2$) parameter is found to be 3.34, which is in the acceptable range. Rietveld refined lattice parameters and atomic positions of constituent elements are listed in table-1 and table-2, respectively. It suggests that the insertion of a layer of SnTe in $Sb_2Te_3$ lattice does not distort the unit cell structure of $Sb_2Te_3$, while the c-axis is enhanced quite appreciably, which is an obvious result. The unit cell of synthesized $SnSb_2Te_4$ single crystal is drawn by using VESTA software and shown in Fig. 2(c). This unit cell contains septuple layers with alternating Sn, Sb, and Te with Sn atoms residing in the middle. These septuple layers are separated from each other through the Vander Waals gap. The unit cell structure of $SnSb_2Te_4$ is different from $Sb_2Te_3$. In $Sb_2Te_3$, the middle atomic layer contains Te(II) atoms [30], while in $SnSb_2Te_4$, the middle atomic layer contains Sn atoms. This middle atomic layer is supposed to directly impact the bulk insulating properties of TIs [31].

The surface morphology of the synthesized $SnSb_2Te_4$ single crystal is visualized through SEM images and shown in figure 3(a). SEM image is showing a typical layered type morphology which signifies laminar growth of the synthesized crystal. It is in well agreement with the XRD pattern recorded on crystal flake which only has reflections of (003n) planes only. Both these results signify the unidirectional growth of the crystal along the c-axis. EDAX mapping of constituent elements viz. Sn, Sb, and Te are shown in Fig. 3(b), 3(c), and 3(d), respectively. It confirms that the elements are distributed homogenously throughout the synthesized crystal. EDAX spectra and elemental composition is shown in Fig. 3(e). EDAX analysis confirms that all constituent elements are present in exact stoichiometric ratios and are homogeneously distributed. No peak for any impurity element can be seen in EDAX spectra, confirming that the synthesized sample is free from any contamination of impurity elements.

Raman spectra are recorded to determine the vibrational modes of synthesized $SnSb_2Te_4$ single crystal. The $SnSb_2Te_4$ shows three sets of vibration modes viz. low-frequency modes ($A^1_{1g}$, $E^1_g$), middle frequency modes ($A_{1g}^2$, $E_g^2$), and higher frequency modes ($A_g^3$ and $E_g^3$) [32]. The vibrational modes denoted by symbol E occur at lower frequency as in these modes out of phase vibrations of atoms of adjacent layers. Conversely, the modes denoted by symbol A arise at a higher frequency, and these modes are formed due to out-of-phase vibrations of atoms of the same atomic layer. Raman modes that occur in $SnSb_2Te_4$ are represented in fig. 4(a). The low-frequency Raman active modes $A_{1g}^1$ and $E_g^1$ consist of out-of-plane vibrations of Sb and Te atoms along the c-axis and in the a-b plane. During these vibrations, the middle Sn layer remains intact. These modes are very similar to the $A_{1g}^1$ and $E_g^1$ modes that occur $Sb_2Te_3$. In these modes, Te atoms vibrate in the same phase as the Sb atom, as shown in fig. 4(a).

Middle frequency modes $A_{1g}^2$ and $E_g^2$ occur due to out of phase vibrations of Te atoms bonded to the middle Sn atomic layer and Sb atomic layer. $A_{1g}^2$ modes occur due to out-of-phase vibrations of Te atoms along the c-axis, and it is the symmetric stretching mode of the bond between Sn and Te. $E_g^2$ modes occur due to out-of-phase vibrations of Te atoms in the a-b plane, and it is the symmetric bending mode of the bond between Sn and Te. These Raman modes strongly depend on the vibrations of Te atoms bonded to the Sn atom. In these vibrations, Te atoms vibrate against Sn and Sb atoms periodically. These modes cannot be observed in the parent compound $Sb_2Te_3$. In the $A_{1g}^2$ and $E_g^2$ modes of $SnSb_2Te_4$, the central part of the lattice takes part in vibrations while in the modes observed in $Sb_2Te_3$, the central part of the lattice remains almost static. As in $Sb_2Te_3$, Raman modes occur due to vibrations of atoms of the outer Te layer while the middle Te layer remains intact [33]. But in middle-frequency Raman modes of $SnSb_2Te_4$, atoms of both internal Te layers take part. It is only possible for $SnSb_2Te_4$ because, in this compound, the most stable middle Wyckoff site is occupied by the Sn atom, which allows atoms of both Te layers to vibrate. In $Sb_2Te_3$, this most stable middle Wyckoff site is occupied by Te atoms that allow the outer Te atoms to vibrate.

The higher frequency Raman modes viz. $A_g^3$ and $E_g^3$ occur due to out-of-phase vibrations of atoms of outer Sb and Te layers. These modes are similar to $A_{1g}^2$ and $E_g^2$ modes that occur in parent compound $Sb_2Te_3$. During these vibrations, the central part of the lattice remains static, as in the case of the $A_{1g}^1$ and $E_g^1$ modes. During $A_g^3$ mode, outer Sb and Te atoms vibrate out of phase along the c-axis, while in $E_g^3$ mode, the out phase vibrations of Sb and Te atoms took place in the a-b plane. Also, in these modes, the Te atoms vibrate out of phase to the Sb atoms, unlike to $A_{1g}^1$ and $E_g^1$ modes, where Te atoms vibrate in phase to the Sb atoms. Due to out of phase moment of Sb and Te, the $A_g^3$ mode is known as the asymmetric stretching mode of Sb and Te, while $E_g^3$ mode is an asymmetric bending mode of Sb and Te.

Fig. 4(b) is showing the recorded Raman spectra of $SnSb_2Te_4$ single crystal. This spectrum is de-convoluted into five peaks by using Lorentz fitting formula. These five peaks are observed at 59.6 cm$^{-1}$, 92.6 cm$^{-1}$, 109.0 cm$^{-1}$, 114.6 cm$^{-1}$ and 162.4 cm$^{-1}$. These modes are identified as $A_{1g}^1$, $E_g^2$, $A_{1g}^2$, $E_g^3$, and $A_g^3$, respectively [32]. These modes are in well agreement with the previous report on Raman active modes of $SnSb_2Te_4$ [32]. Low-frequency $E_g^1$ mode could not be detected here as the spectra are recorded above 50cm$^{-1}$, and this mode occurs well below 50cm$^{-1}$. Detection of $E_g^2$ and $A_{1g}^2$ modes in Raman Spectra confirms that the layer SnTe has been successfully inserted in the $Sb_2Te_3$ lattice as these modes strongly depend on the bond between the middle Sn and Te atom.

Results of magneto-transport measurements of $SnSb_2Te_4$ are shown in Fig. 5(a). The inset of fig. 5(a) is showing normalized resistivity vs. temperature measurements plot from 250K down to 5K. All resistivity values are normalized. It is clear from this plot that the synthesized $SnSb_2Te_4$ single crystal is metallic in nature as the resistivity values are decreasing with lowering the temperature. Fig. 5(a) shows variation in magneto-resistance percentage (MR%) of synthesized $SnSb_2Te_4$ single-crystal w.r.t. applied magnetic field at 5K. MR% is calculated by using the following formula

$$MR\% = [\rho(H) - \rho(0)] * 100 / \rho(0)$$

Here, $\rho(H)$ represents resistivity in an applied magnetic field, and $\rho(0)$ represents resistivity in the absence of an applied magnetic field or at zero fields. Here MR% data is taken in both directions, and the mean is calculated to uphold the symmetry of the plot. In MR% measurements, $SnSb_2Te_4$ exhibits a non-saturating MR that reaches a meager value of 0.125% at 5K under a magnetic field in a range from +5T to -5T. Interestingly, at low field up to +-1T, MR% have a sharp V-like shape. This type of behavior of MR% at the low field is the signature of the presence of the WAL effect in the measured sample [34,35]. It suggests that back-scattering is suppressed at a low magnetic field due to π- Berry phase of Dirac fermions existing in surface states [34-36]. This V-like shape in MR at a low field signifies that the synthesized $SnSb_2Te_4$ sample shows a WAL effect at low temperature. It gives magneto-transport evidence of non-trivial topological character and the presence of surface states in $SnSb_2Te_4$ single crystal.

Figure 5(b) shows the variation of magneto-conductivity vs. applied magnetic field in a range of ±1T at temperature 5K for $SnSb_2Te_4$ crystal. In topological materials, the physical parameters which characterize the weak anti-localization effect have been calculated by using

Hikami-Larkin-Nagaoka (HLN) model [37]. Here, Δσ(H) is given as the difference between conductivity at the applied field (σ(H)) and zero fields (σ(0)). According to the HLN model, the magneto-conductivity can be described as

$$\Delta\sigma(H) = -\frac{\alpha e^2}{\pi h}\left[\ln\left(\frac{B_\varphi}{H}\right) - \Psi\left(\frac{1}{2} + \frac{B_\varphi}{H}\right)\right]$$

where $B_\varphi = \frac{h}{8e\pi L_\varphi^2}$ is the characteristic field, $L_\varphi$ is phase coherence length, Ψ is digamma function, e is the electronic charge, h is Plank's constant, and H is applied magnetic field. The prefactor α takes the value -0.5 per conduction channel, and $L_\varphi$ is the distance traveled by the electron up to which it remains its phase. Also, the pre-factor α characterizes the type of localization present in the material. In fig. 5(b), the obtained magneto-conductivity is fitted with the HLN equation in low field regime (up to ±1T), represented by the red curve. The extracted values of fitting parameters α and $L_\varphi$ are -1.249×10$^{-5}$ and 61.3217 nm, respectively. The obtained α value indicates the presence of a weak anti-localization effect and 2D conduction. The observed value of α is very low, which is directly related to the number of non-trivial topological states. The standard value for a single topological conducting channel is 0.5. The lower or higher values of α from 0.5 suggest that the conductivity contributes to other states [38,39]. Here, the observed value of α is much smaller than the standard value of the same; this indicates that topologically trivial states also contribute to the conductivity and non-trivial states. This lower value of a fitting factor α is consistent with some previous reports on materials showing low MR% [40,41]. Overall, it can be summarized that in addition to surface states, there is a contribution of bulk states as well in overall conduction in SnSb$_2$Te$_4$ crystal.

Fig. 6(a) shows the calculated DOS of synthesized SnSb$_2$Te$_4$ single crystal within the protocols of Density Functional Theory (DFT). Rietveld refinement crystal parameters are considered to calculate DOS and the band structure theoretically. These calculations measure the spin-orbit coupling (SOC) and without SOC effects as implemented in Quantum Espresso with Perdew-Burke-Ernzerhof (PBE) exchange-correlation functional [42,43]. The right-hand side image shows the calculated DOS of SnSb2Te4 without SOC while the left hand side image is showing shows with inclusion of SOC. These figures show a uniform spread of DOS in energy range -2eV to 2eV, suggesting that there is covalent bonding between atoms in SnSb$_2$Te$_4$. Projected DOS are also calculated to determine the contributions of Orbitals separately. In projected DOS, p orbitals of Sn, Sb and Te are found to be the major contributors. DOS is significantly decreased at the Fermi level, as a sharp dip is observed in the DOS plot at the Fermi level. This dip is clearly observable in both without SOC and with SOC, DOS plots. The dip in DOS at the Fermi level is different in $SnSb_2Te_4$ as compared to the parent compound $Sb_2Te_3$. In $Sb_2Te_3$, DOS is wholly vanished at the Fermi level, signifying the bulk insulating property of Sb2Te3, and hence Sb2Te3 is known as a topological insulator for $Bi_2Te_3$, which is also a topological insulator [44]. Here in $SnSb_2Te_4$, DOS is decreased significantly but is not entirely vanished. It suggests that the bulk of $SnSb_2Te_4$ is not completely insulating. These non-vanishing DOS at the Fermi level suggest that $SnSb_2Te_4$ can be regarded as a topological metal or semimetal. This metallic behavior observed in DOS calculations agrees with the metallic behavior observed in ρ-T measurements, as shown in the inset of Fig. 5(a). The same feature in DOS is also observed in $GeBi_2Te_4$ [45], in which a layer of GeTe is inserted into the lattice of $Bi_2Te_3$; this was also considered a topological metal.

Fig. 6(b) is showing the calculated bulk electronic band structure without SOC and with SOC. These calculations are performed through the K-path $S_0 \rightarrow \Gamma \rightarrow L \rightarrow H_0$, calculated from the SeeK-path: the k-pathfinder and visualizer [46]. The Left-hand side plot of Fig. 6(b) is showing the calculated band structure without SOC, while the right-hand side plot is showing the same with SOC. The electronic band structure without SOC is completely gapped at Fermi level while there is the crossing of bands at Fermi level in with SOC plots. A similar feature of electronic band structure was observed for other topological metal candidate $GeBi_2Te_4$ [45]; bulk electronic bands were completely gapped at Γ point while a Dirac cone was observed below Fermi level with the inclusion of SOC.

Right hand side image of Fig. 6(b) shows bulk electronic band structure with the inclusion of SOC. A significant impact of SOC is evident as the band is inverted when SOC is switched ON. In bulk electronic band structure with SOC shown in Fig. 6(b), Γ the point is the high symmetry point of the Brillouin zone, at which the impact of SOC can be clearly seen. Bulk electronic band structure around Γ point with and without SOC is shown in Fig. 6(c). All bands at the Γ point are inverted and showing the anti-crossing features with the inclusion of SOC, and this signifies that the SOC has a significant impact on the electronic band structure. A Dirac point is also observed in with the SOC plot at Γ point which lies at 0.016eV energy below the Fermi level. This Dirac point is shown in the right-hand side image of Fig. 6(c). This effective SOC and non-vanishing DOS at Fermi level suggest that SnSb2Te4 cab ne is regarded as a new member of family of Topological metals and similar compound $GeBi_2Te_4$.

**Conclusion:**

Summarily, a single crystal of $SnSb_2Te_4$ is grown by using a simple self-flux method. Crystalline growth and phase purity are evident from XRD, SEM, and EDAX measurements. Two different Raman modes are observed in parent Sb2Te3 due to changes in the middle atomic layer. It further signifies that the synthesized crystal has the phase of $SnSb_2Te_4$. Metallic behavior is evident from the resistivity vs. temperature plot. Here, we are the first to report MR% behavior of $SnSb_2Te_4$ at 5K under the magnetic field in a range of -5T to +5T. Despite low MR%, observation of V-like cusp at low field signifies WAL effect in synthesized $SnSb_2Te_4$ single crystal. This WAL effect is also confirmed through HLN fitting, indicating a contribution of bulk states along with the topological surface states in the conductivity of $SnSb_2Te_4$. DFT calculations also signify that SOC is effective on the band structure of $SnSb_2Te_4$ and suggest this material to be topological metal or semimetal in contrast to its parent compound $Sb_2Te_3$. Altogether, this is the first-ever report on the presence of WAL in SnSb2Te4 which can be regarded as the first report on experimental evidence of the topological character of $SnSb_2Te_4$. This report will undoubtedly open new doors to explore this system in the context of the topological behavior of this material.

**Acknowledgment:**

The authors would like to thank Director NPL for his keen interest and encouragement. The authors are thankful to Mr. Krishna Kandpal for the vacuum encapsulation of the sample. Ankush Saxena would like to thank DST for the research fellowship. M.M. Sharma and Yogesh Kumar would like to thank CSIR for the research fellowship. Prince Sharma would like to thank UGC for the research fellowship. Ankush Saxena, M.M. Sharma, Prince Sharma, and Yogesh Kumar are also thankful to AcSIR for Ph.D. registration.

**Author statement:**

All authors are equally contributing to this article.

**Table-1.**

Unit cell parameters obtained from Rietveld refinement of PXRD pattern of synthesized SnSb$_2$Te$_4$ single crystal:

| Cell Parameters | SnSb$_2$Te$_4$ |
|---|---|
| Structure | Rhombohedral |
| Space Group | R -3 m |
| a | 4.4034(2) |
| b | 4.4034(2) |
| c | 41.6287(2) |
| α | 90 |
| β | 90 |
| γ | 120 |

**Table-2**

Atomic positions of constituent elements of synthesized SnSb$_2$Te$_4$ single crystal:

| Atom | x | y | z |
|---|---|---|---|
| Sn | 0.0000 | 0.0000 | 0.0000 |
| Sb | 0.0000 | 0.0000 | 0.42729 |
| Te1 | 0.0000 | 0.0000 | 0.14197 |
| Te2 | 0.0000 | 0.0000 | 0.28556 |

**Figure captions:**

Fig. 1: Schematic of heat treatment followed to synthesize $SnSb_2Te_4$ single crystal nd inset is showing the image of synthesized $SnSb_2Te_4$ single crystal.

Fig. 2(a): XRD pattern taken on mechanically cleaved crystal flake of synthesized $SnSb_2Te_4$ single crystal.

Fig. 2(b): Rietveld refined PXRD pattern of synthesized $SnSb_2Te_4$ single crystal.

Fig. 2(c): Unit cell of synthesized $SnSb_2Te_4$ single crystal by using VESTA software.

Fig. 3(a): SEM image of synthesized $SnSb_2Te_4$ single crystal (b) EDAX mapping of $SnSb_2Te_4$ showing distribution of Sn (c) EDAX mapping of $SnSb_2Te_4$ showing distribution of Sb (d) EDAX mapping of $SnSb_2Te_4$ showing distribution of Te (e) EDAX spectra showing elemental composition of synthesized $SnSb_2Te_4$ single crystal.

Fig. 4(a): Illustration of Raman modes that occur in $SnSb_2Te_4$.

Fig. 4(b): De-convoluted Raman spectrograph of $SnSb_2Te_4$ at room temperature.

Fig. 5(a): MR% vs applied field plot of synthesized $SnSb_2Te_4$ single crystal at 5K under the applied magnetic field in a range of -5T to +5T, inset is showing normalized resistivity vs temperature plot from 250K to 5K.

Fig. 5(b): HLN fitted conductivity plot in low magnetic field range ±1T at 5K.

Fig. 6(a): DFT calculated DOS along with projected DOS of $SnSb_2Te_4$ with and without SOC.

Fig. 6(b): Bulk electronic band structure of $SnSb_2Te_4$ calculated under DFT protocols with and without SOC.

Fig. 6(c): Bulk electronic band structure of $SnSb_2Te_4$ at Γ point with and without SOC, right hand size image is showing the zoomed view of Dirac point occurring below Fermi level at Γ point.

Fig. 1

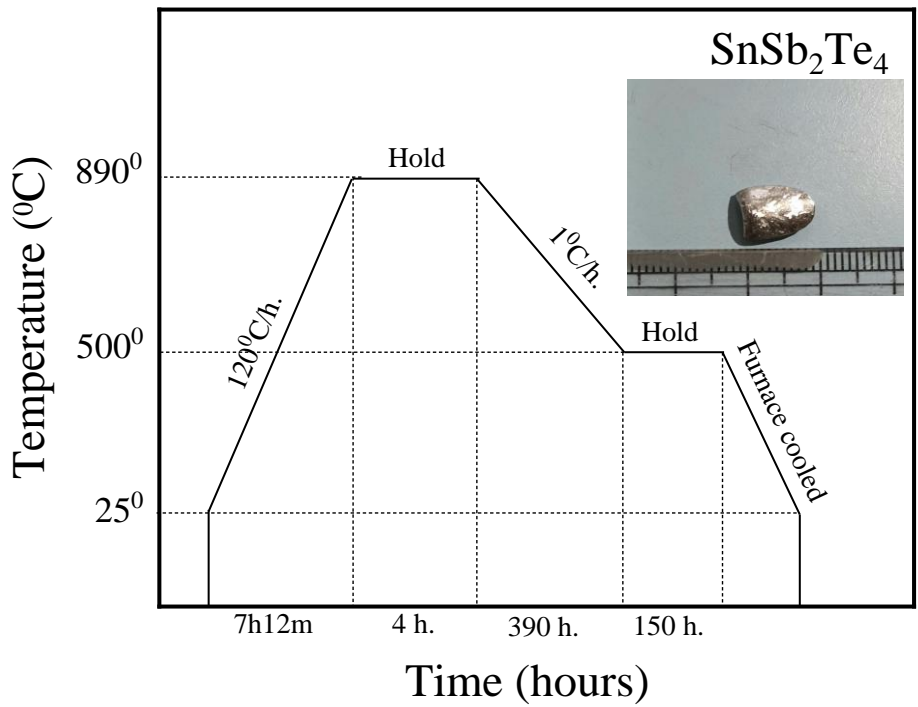

Fig. 2(a)

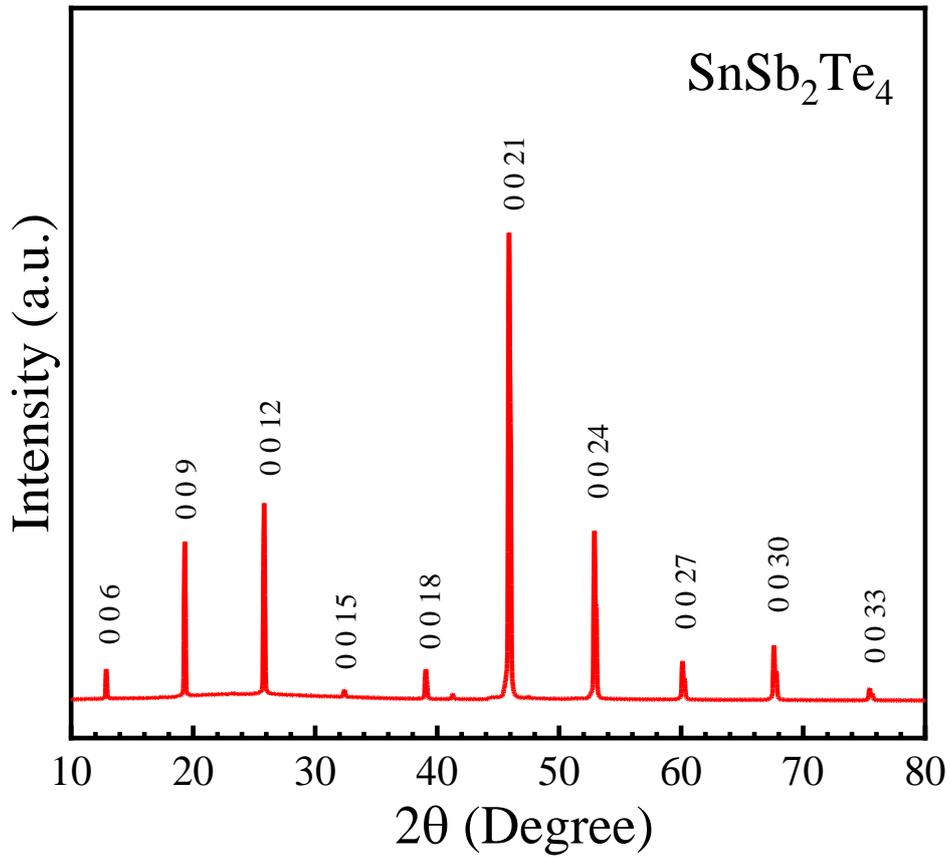

Fig. 2(b)

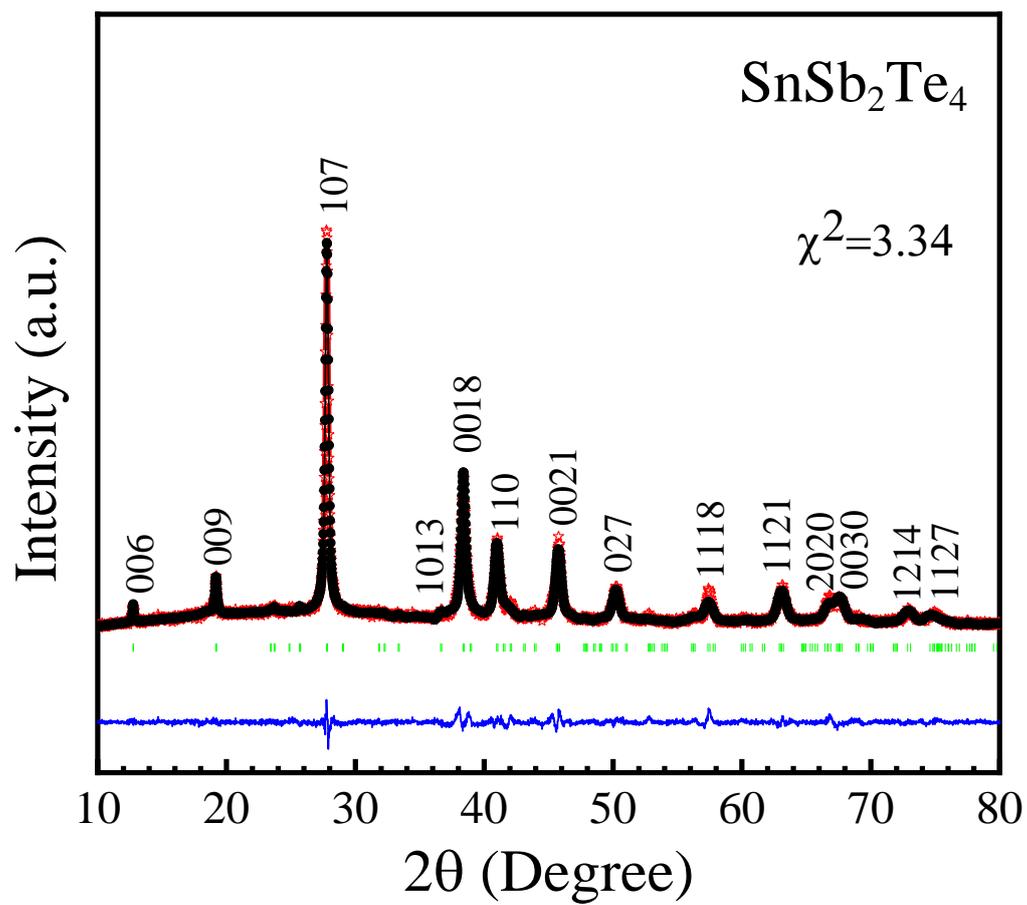

Fig. 2(c)

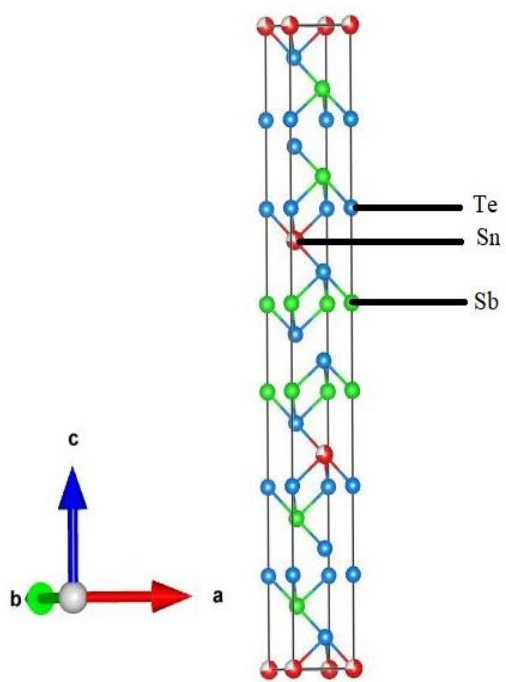

Fig. 3

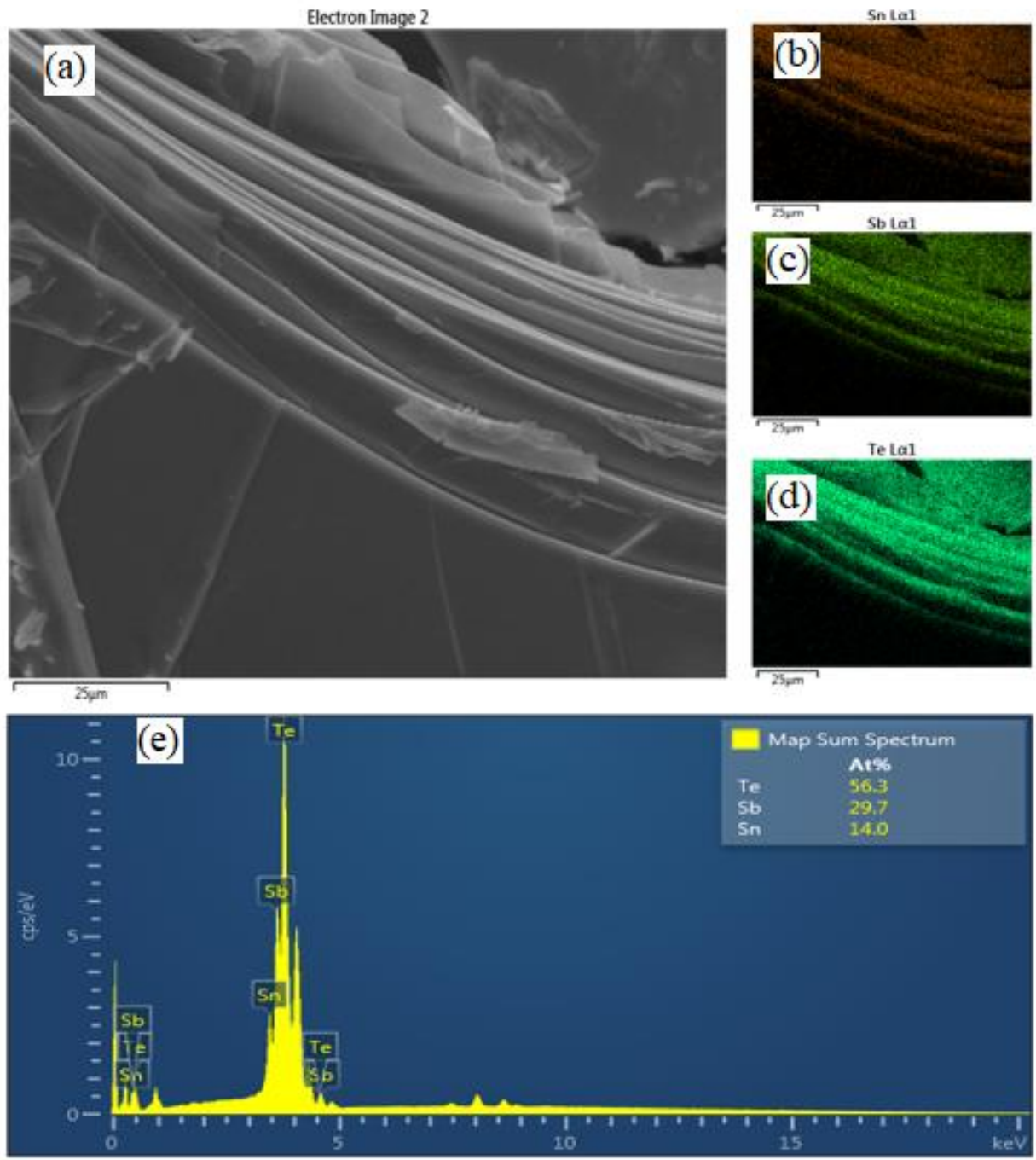

Fig. 4(a)

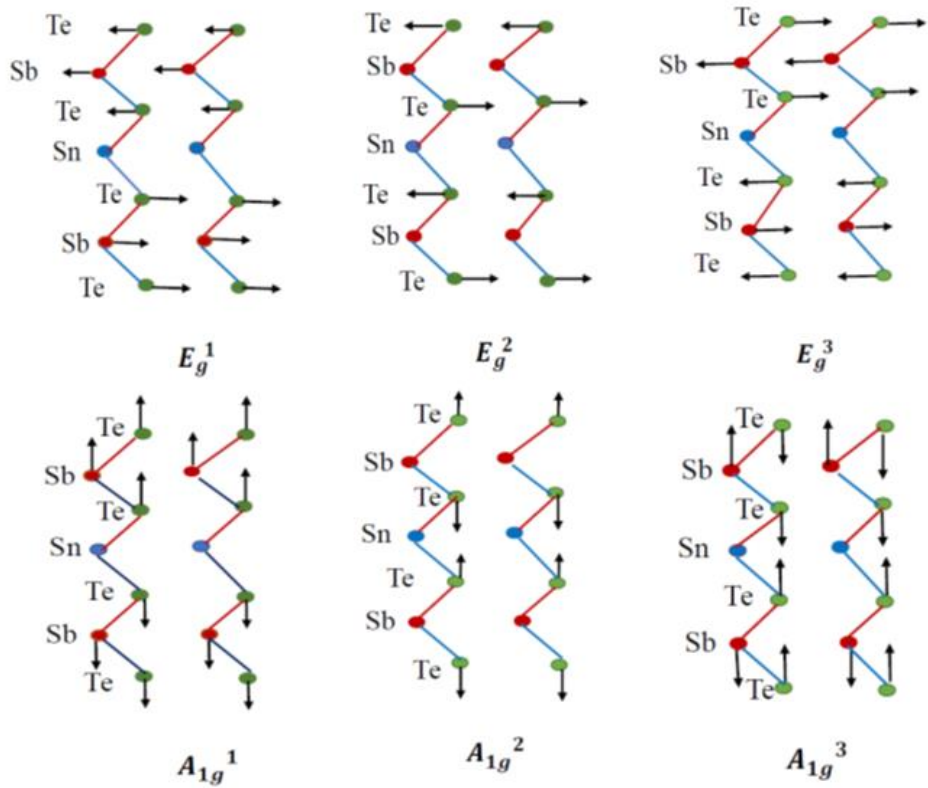

Fig. 4(b)

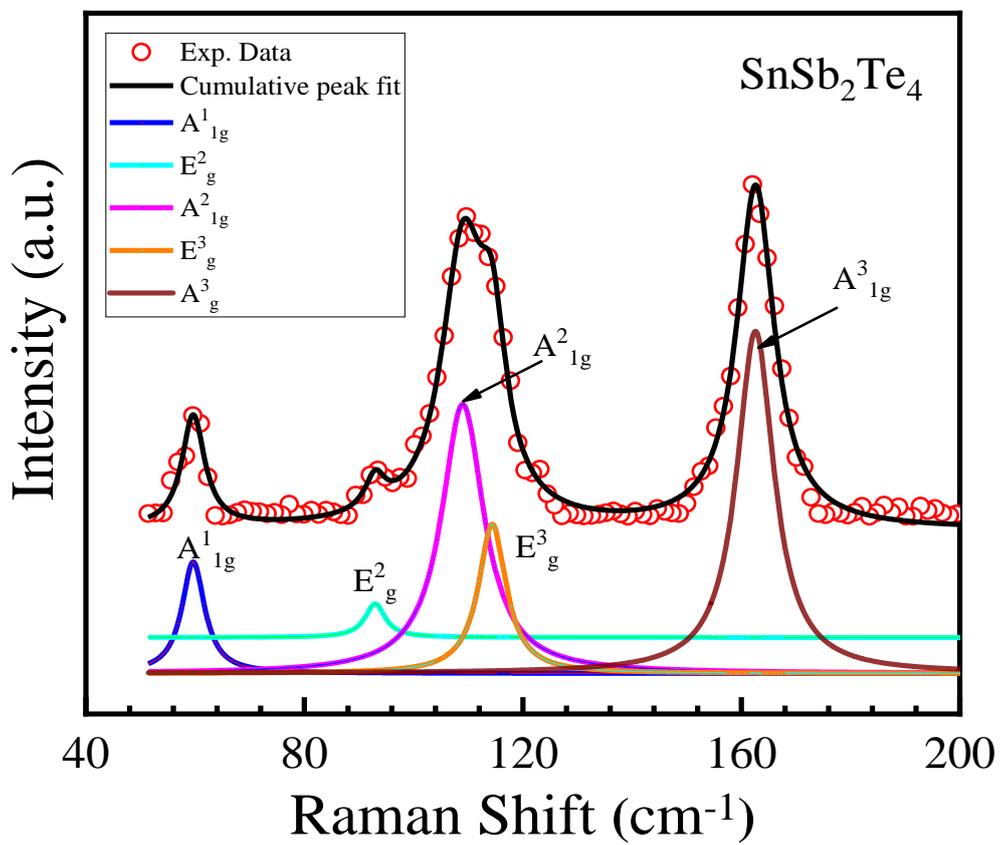

Fig. 5(a)

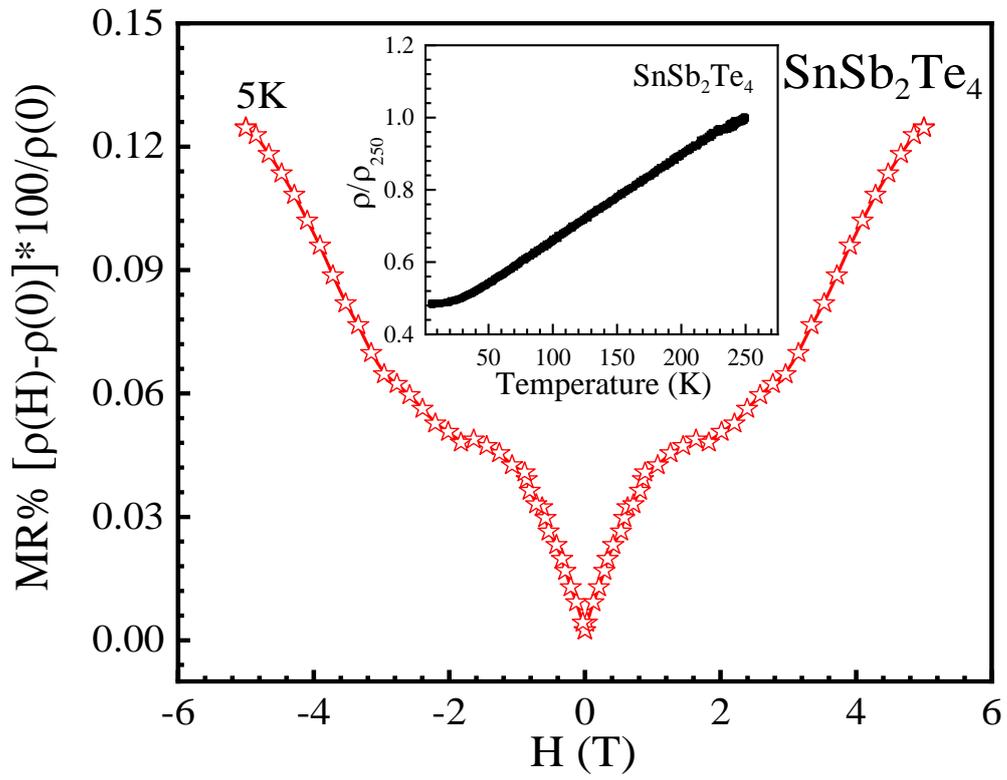

Fig. 5(b)

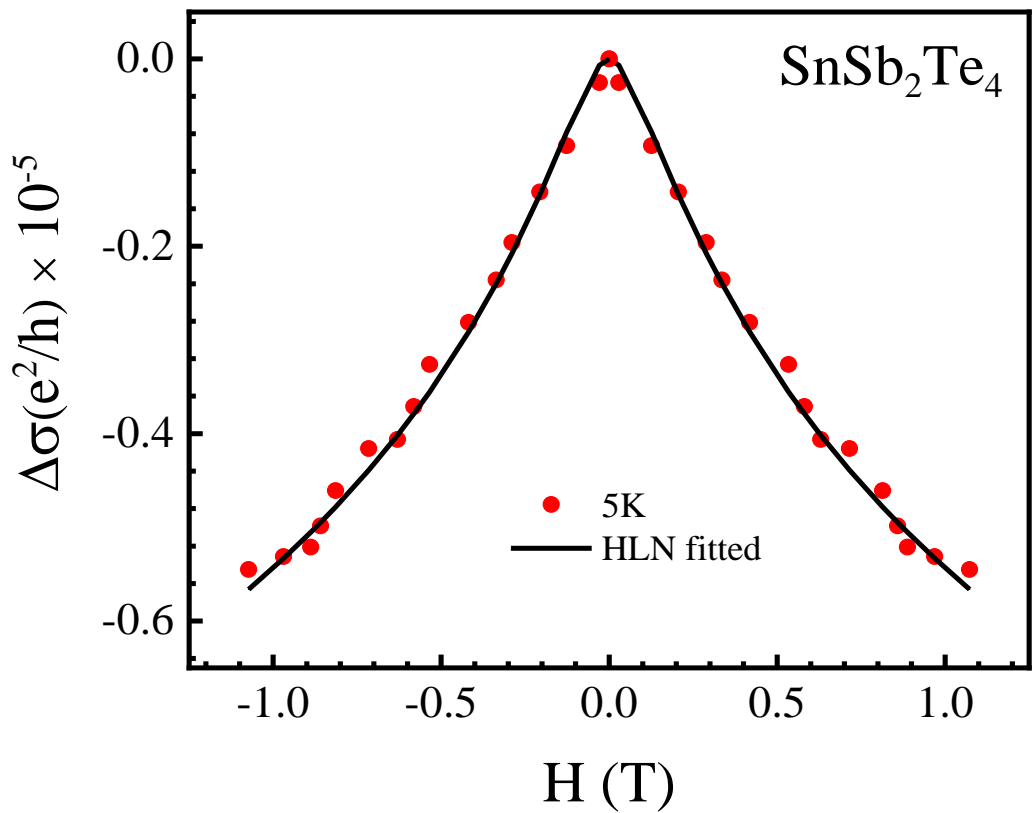

Fig. 6(a)

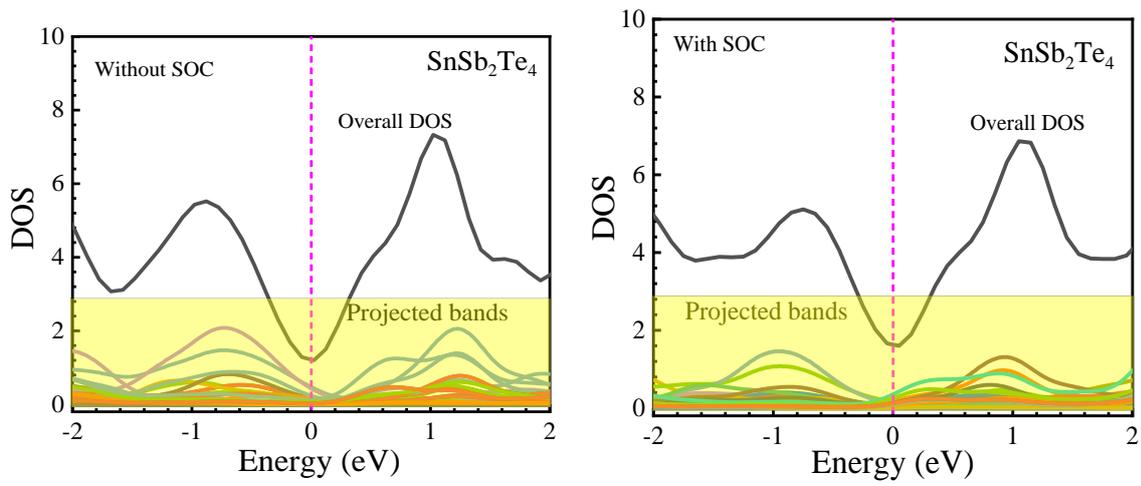

Fig. 6(b)

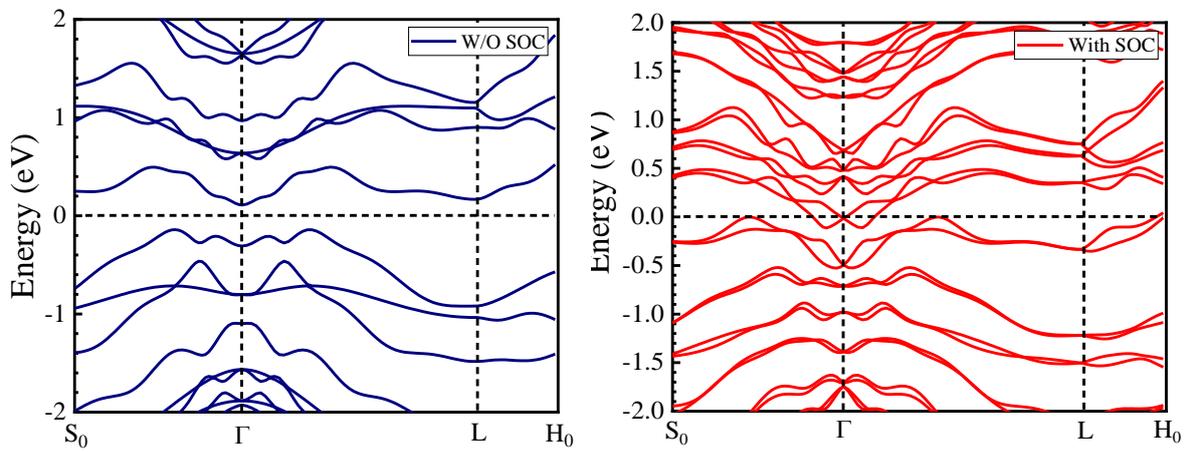

Fig. 6(c)

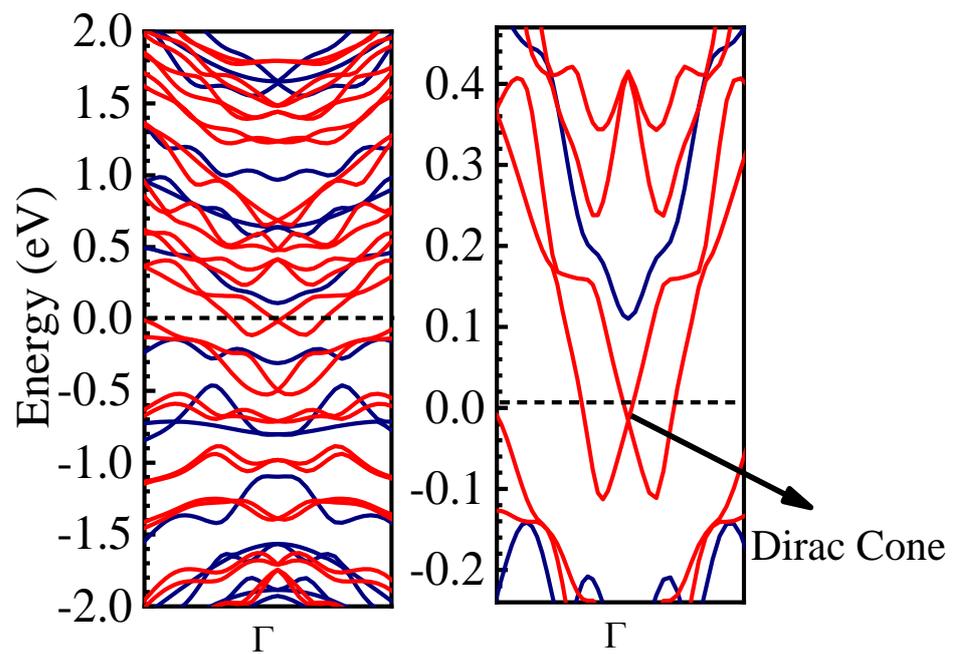